\documentclass[reprint,prl,nofootinbib]{revtex4-1}
\pdfoutput=1
\usepackage{graphicx,amsmath}
\newcommand{\eq}[1]{\begin{equation}#1\end{equation}}
\newcommand{\eqa}[1]{\begin{eqnarray}#1\end{eqnarray}}
\newcommand{\secs}[1]{\section{#1\label{sec-#1}}}

\newcommand{\fig}[4]{\begin{figure}[#4]\centering\includegraphics[width=#3\textwidth]{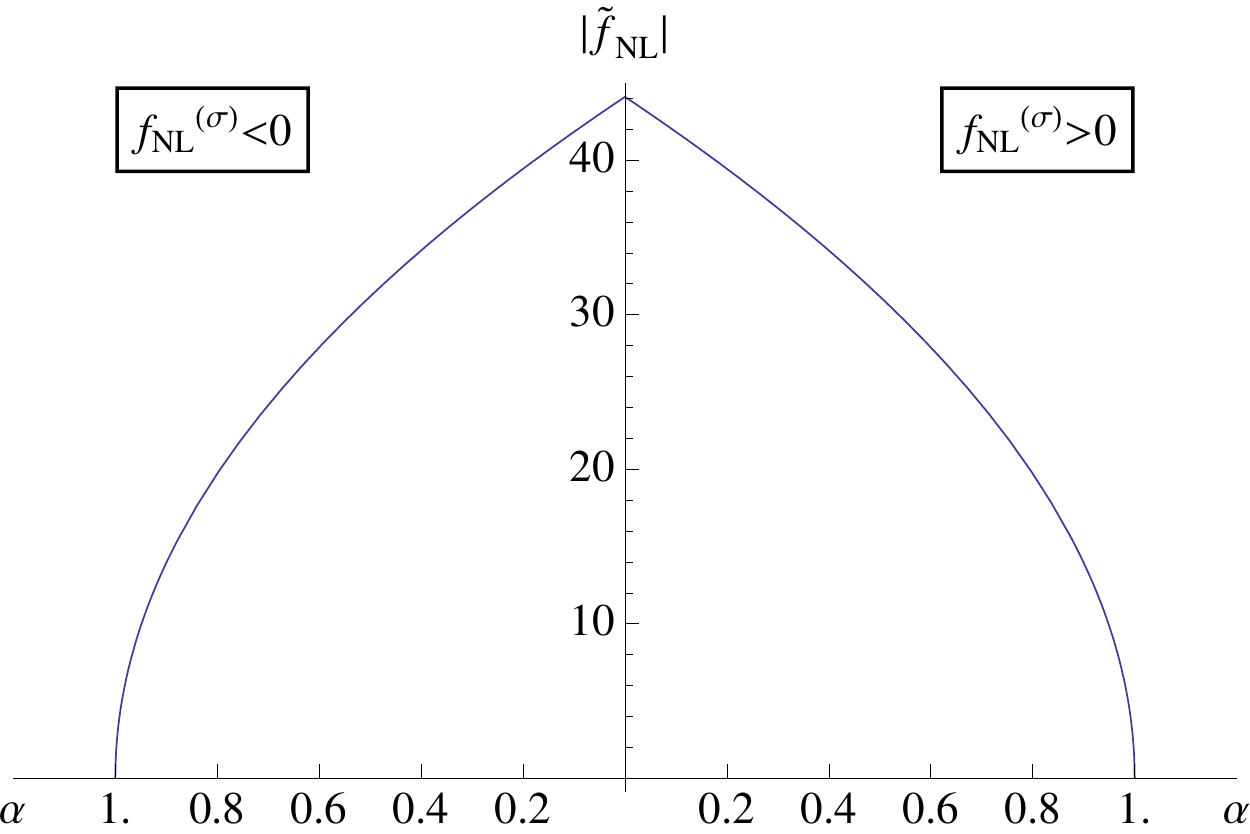}\caption{#2}\label{fig-fNL}\end{figure}}

\newcommand{\refeq}[1]{Eq.\ (\ref{eq-#1})}

\newcommand{\refig}[1]{FIG.\ \ref{fig-#1}}

\newcommand{\subs}[1]{_\mathrm{#1}}
\newcommand{\sups}[1]{^\mathrm{#1}}
\newcommand{\dd}[1]{\mathrm{d}#1}

\def\fNL{f\subs{NL}{}}
\def\tNL{\tau\subs{NL}{}}

\begin{document}
\title{Small non-Gaussianity and dipole asymmetry in the CMB}
\author{Lingfei Wang}
\author{Anupam Mazumdar}
\affiliation{Consortium for Fundamental Physics, Lancaster University, Lancaster LA1 4YB, UK}
\begin{abstract}
In this paper we provide a prescription for obtaining a small non-Gaussianity and the observed dipole asymmetry in the cosmic microwave background radiation. The observations inevitably lead to multi-field inflationary dynamics, where each field can create positive or negative large non-Gaussianity, resulting a fine cancellation but with an observable imprint on the hemispherical asymmetry. We discuss this possibility within a simple slow-roll scenario and find that it is hard to explain the observed dipole asymmetry. We briefly discuss some speculative scenarios where one can explain dipole asymmetry.
\end{abstract}
\maketitle

\section{Introduction}
The current observations from the Planck suggest that the observed non-Gaussianity is 
very tiny~\cite{Planck-NG}, and the limits are all consistent with a single {\it canonical} field inflation model with Bunch-Davis initial 
vacuum condition~\cite{Planck-infl, Visible}~\footnote{The data is also compatible with uncoupled multi-field models of inflation provided all the fields have 
a unique late-time attractor behavior giving rise to solely {\it adiabatic } perturbations, such as in the case of {\it Assisted Inflation}~\cite{Assist}.}. On the 
other hand it has also been observed that 
there exists a hemispherical asymmetry 
in the cosmic microwave background (CMB) radiation~\cite{Planck-anisotropy}, confirming the earlier results of WMAP~\cite{Wmap-ani,Chris,Prunet}.
It has been argued that initial fluctuations in one of the fields other than the inflaton~\cite{Erickcek:2008sm,Dai:2013kfa} can {\it in principle} imprint this asymmetry provided this hemispherical asymmetry is modeled by the dipolar modulation in the CMB anisotropy: ${\cal P}_{\zeta}(k,r)= {\cal P}_{\zeta}(k)[1+2A\vec p\cdot\vec r/r_{ls}]$, where $r_{ls}$ is the distance to the last scattering surface. For the present comoving modes, $ {\cal P}_{\zeta}(k)$ is the isotropic Gaussian perturbations, $A$ is the magnitude of the dipole asymmetry and its direction is given by $\vec r$~\cite{Chris}. The amplitude is bounded by 
$A=0.072\pm 0.022$ for $\ell \leq 64$~\cite{Planck-anisotropy,Wmap-ani}. The challenge is to explain this anisotropy without altering the large scale homogeneity
and the amplitude of the density perturbations~\footnote{By the existence of initial super-Hubble perturbations, we mean that the perturbations in a particular field are already present even before the pivot scale $k_\ast$.}. 

One particular explanation arises  via the mode-mode interactions, where small and long wavelength modes can couple. This can arise naturally if the
perturbations are non-Gaussian. In the case of a single canonical field responsible for creating the curvature perturbations, the strength of the dipole anisotropy would be proportional to the local bispectrum, i.e. $|A| \propto \fNL$, see for instance, Ref.~\cite{Erickcek:2008sm,Dai:2013kfa,Lyth:2013vha}. However, when there are multiple sources co-exist, it is 
possible to create a {\it positive} and a {\it negative} local bispectra, i.e. $\pm \fNL$, from different perturbation sources, where they partly cancel and leave us with a smaller total bispectrum, i.e. $\fNL\sim{\cal O}(1)$. 
Typically one expects $+\fNL$, but $-\fNL$ can be obtained in many scenarios, such as non-Gaussianity generated during preheating~\cite{Enqvist:2004ey,Jokinen:2005by}, or in a spectator scenario where the spectator field seeds the perturbations during inflation but decays before the end of inflation~\cite{Mazumdar:2012rs}. 

\section{Fine cancellation between $\pm\fNL$ }
In order to illustrate our point, let us consider a very simple setup with two fields, $\sigma$ and $\psi$,  dominantly sourcing the curvature perturbations.
 According to $\delta N$ formalism, see for instance~\cite{Stewart}, the number of e-folds of universe expansion, $N$, can be written as a function of the fields $N(\sigma,\psi)$. Any perturbation $(\delta\sigma,\delta\psi)$ in the fields  would generate the perturbation (up to the second order)
\eq{\delta N=N_\sigma\delta\sigma+N_\psi\delta\psi+\frac{1}{2}\Bigl(N_{\sigma\sigma}\delta\sigma^2+N_{\sigma\psi}\delta\sigma\delta\psi+N_{\psi\psi}\delta\psi^2\Bigr),}
where the subscripts mean derivatives w.r.t  the fields $\psi$ and $\sigma$. The power spectrum of the curvature perturbations then becomes
\eq{P_\zeta=P_{\delta N}=N_\sigma^2P_{\delta\sigma}+N_\psi^2P_{\delta\psi}=\Bigl(N_\sigma^2+N_\psi^2\Bigr)\frac{H_*^2}{4\pi^2},\label{eq-E-Pz0}}
where the last equivalence comes from the quantization of the slow roll fields in the inflationary background, and $H_*$ is the Hubble rate of expansion during inflation.
 
The primordial local bispectrum is then given  by, see \cite{Lyth:2005fi}:
\eq{\fNL=\frac{5}{6}\frac{N_{\sigma\sigma}N_\sigma^2+N_{\psi\psi}N_\psi^2+2N_{\sigma\psi}N_\sigma N_\psi}{(N_\sigma^2+N_\psi^2)^2}.\label{eq-E-fNL0}}
Let us define $\alpha$ as the contribution to the curvature perturbations arising from the $\sigma$ field, 
\eq{\alpha\equiv\frac{N_\sigma^2}{N_\sigma^2+N_\psi^2},\hspace{0.5in}0<\alpha<1.}
In this part of the calculation, we are interested in the simplest case where the last term in \refeq{E-fNL0} (i.e.\ $N_{\sigma\psi}$) is negligible. This can be satisfied in many non-interacting models of inflation, see for instance~\cite{Infl-rev}. In such scenarios, the primordial local bispectrum can be written as
\eq{\fNL=\alpha^2\fNL\sups{(\sigma)}+(1-\alpha)^2\fNL\sups{(\psi)}.\label{eq-E-fNL1}}
Here the individual bispectra are defined as
\eq{\fNL^{(\sigma)}\equiv\frac{5}{6}\frac{N_{\sigma\sigma}}{N_{\sigma}^2},\hspace{0.35in}\fNL^{(\psi)}\equiv\frac{5}{6}\frac{N_{\psi\psi}}{N_{\psi}^2}.}

In general, there is a possibility in \refeq{E-fNL1} to get opposite significant contributions from the two terms, so they mostly cancel and only generate a small total $\fNL$ under our current observational limit~\cite{Planck-NG}. For example, they can both be ${\cal O}(30)$, but with opposite signs. Their amplitudes cannot be arbitrarily large on the other hand, because they are constrained by the observational limit on the trispectrum parameter
\eqa{\tNL&=&\frac{N_\sigma^2N_{\sigma\sigma}^2+N_\psi^2N_{\psi\psi}^2}{(N_\sigma^2+N_\psi^2)^3}\nonumber\\
&=&\frac{36}{25} \left(\alpha^3(\fNL\sups{(\sigma)})^2+(1-\alpha)^3(\fNL\sups{(\psi)})^2\right),\label{eq-E-tNL0}}
where we have also neglected the $N_{\sigma\psi}$ contribution.

\secs{Initial fluctuation from a single field}
When an initial fluctuation exists, it can lead to different field configurations at the Hubble exit of the pivot scale in different Hubble patches, and therefore a modulation in the power spectrum will arise naturally~\cite{Erickcek:2008sm}. Let us consider two separate Hubble patches with an initial fluctuation $\Delta\sigma$ in the $\sigma$ field alone. It will lead to a modulation in the power spectrum of the curvature perturbations, indicated by the parameter, see~\cite{Erickcek:2008sm}
\footnote{The conventions of $A$ are different in \cite{Planck-anisotropy} and \cite{Erickcek:2008sm}, by a factor of $4$. We are using the convention of the Planck satellite in \cite{Planck-anisotropy}, so our \refeq{E-A0} from \cite{Erickcek:2008sm} acquires an additional coefficient $4$.}
\eq{A=\frac{\Delta P_\zeta}{4P_\zeta}=\frac{1}{4P_\zeta}\frac{\partial P_\zeta}{\partial\sigma}\Delta\sigma.\label{eq-E-A0}}

In our simplest setup, the perturbation $\Delta\sigma$ will not cause any non-negligible perturbation in the total energy 
density, or $H_*$. This means
\eq{|A|=\left|\frac{N_\sigma N_{\sigma\sigma}}{2(N_{\sigma}^2+N_\psi^2)}\Delta\sigma\right|=\frac{3}{5}\frac{|\Delta\sigma|}{\sqrt{P_{\delta\sigma_*}}}\sqrt{P_\zeta}\,\left|\widetilde\fNL\right|,\label{eq-E-A1}}
where we have defined the \emph{effective} $\fNL$ as
\eq{\widetilde\fNL\equiv\alpha^\frac{3}{2}\fNL\sups{(\sigma)}.}

Now,  \refeq{E-A1}, can be seen as  the dipolar asymmetry $A$ being proportional to the effective $\widetilde \fNL$, which corresponds to the observed $\fNL$ when only one field contributes to the curvature perturbations. In the limit $\alpha\rightarrow1$, i.e.\ when $\sigma$ totally dominates the curvature perturbations, we get $\widetilde\fNL\rightarrow\fNL$, as in \cite{Lyth:2013vha}.

However if we allow a deviation from $\alpha\rightarrow1$, it is possible to achieve a much larger $\widetilde \fNL$, and therefore an enhanced CMB asymmetry $|A|$, within a small observed $\fNL$. 
This can be realized by a (partial or complete) cancellation in \refeq{E-fNL1}, between the opposite contributions by the two fields $\sigma$ and $\psi$.

Since the cancellation in \refeq{E-fNL1} is constrained by $\tNL$ in \refeq{E-tNL0}, we can solve $\fNL\sups{(\sigma)}$ and $\fNL\sups{(\psi)}$ from \refeq{E-fNL1} and \refeq{E-tNL0}. This allows us to write the effective $\widetilde\fNL$ as a function of the observed $\fNL$ and $\tNL$, as
\eqa{\mathrm{for\ }&&\pm\fNL\sups{(\sigma)}>0,\nonumber\\
&&\widetilde\fNL=\sqrt\alpha\fNL\pm\sqrt{1-\alpha}\sqrt{\frac{25}{36}\tNL-\fNL^2}.}

The latest observation constrains the primordial bispectrum and trispectrum by $-8.9<\fNL<14.3$ and $\tNL<2800$ at $95\%$ CL.~\cite{Planck-NG}. As an example, we take $\fNL=0$, $\tNL=2800$, and find $|\widetilde\fNL|$ can reach a maximum of 45 when $\alpha\rightarrow0$. This is shown in \refig{fNL}.
\fig{fNL}{The effective $|\widetilde\fNL|$ as a function of $\alpha$, in the case of the observed $\fNL=0$ and $\tNL=2800$.}{0.5}{}

\secs{Initial fluctuations from multi-field}
We now turn to the general multi-field inflationary models, with $\phi_\mu$, $\mu=0,1,\dots,n-1$, as all the existing $n$ fields collectively responsible for inflation and perturbations. We assume they are all canonical slow-roll fields, so according to $\delta N$ formalism the primordial local trispectrum parameter is given by:
\eq{\tNL=\frac{N_\mu N_{\mu\lambda}N_{\lambda\nu}N_\nu}{\Bigl(\sum\limits_\mu N_\mu^2\Bigr)^3}=\frac{\sum\limits_\mu Q_\mu^2}{\Bigl(\sum\limits_\mu N_\mu^2\Bigr)^3}.\label{eq-M-tNL0}}
Here we do not neglect the cross-coupling terms $N_{\mu\nu}$, where $\mu\ne\nu$, and we have defined~\footnote{It is worth noting that in two field scenarios, $Q_\mu$, or the second order dynamics can be exactly solved from the expressions of spectral index $n_s$ and the local bispectrum $\fNL$, and hence also the asymmetry parameter, $A$. Then $A$ is solely determined by the background dynamics of the universe and $\alpha$, giving precise values for $A$ instead of the inequality in \refeq{M-Aa1}.}
\eq{Q_\mu\equiv N_\nu N_{\nu\mu}.}

Now, if some initial fluctuation generates a field difference $\Delta\phi_0$ for $\phi_0$ at the Hubble exit of pivot scale, while leaving other fields $\phi_i$, $i=1,2,\dots,n-1$ unperturbed, it is straightforward to show that the CMB anisotropy would be~\footnote{\label{fn-multi}In general, the field differences can occur for all the fields as $\widetilde{\Delta\phi_\mu}$. However,  for canonical fields we can always perform a rotational field redefinition, so any field difference configuration is aligned along the $\phi_0$ direction. After the rotation we will get $\Delta\phi_0=\sqrt{\sum\limits_\mu\widetilde{\Delta\phi_\mu}^2}$, but $\Delta\phi_i=0$. This allows us to perform the follow-up analysis.}
\eq{A=\frac{Q_0}{2\sum\limits_\mu N_\mu^2}\Delta\phi_0,}
by assuming that the change in the total energy density by $\Delta\phi_0$ will not create significant dipole asymmetry $A$.

From \refeq{M-tNL0}, we know that
\eq{Q_0^2<\sum\limits_\mu Q_\mu^2=\tNL\Bigl(\sum\limits_\mu N_\mu^2\Bigr)^3.}
This gives an upper bound for
\eq{|A|<\frac{1}{2}|\Delta\phi_0|\sqrt{\tNL\sum\limits_\mu N_\mu^2}=\frac{|\Delta\phi_0|}{2\sqrt{P_{\delta\phi_{0*}}}}\sqrt{\tNL P_\zeta},\label{eq-M-Aa1}}
where the power spectrum of the pivot scale field perturbations $\delta\phi_{0*}$ is $P_{\delta\phi_{0*}}=H_*^2/4\pi^2$ for canonical slow roll fields.

By substituting the recent observational value of $P_\zeta=2.196\times10^{-9}$, and the latest upper bound on $\tNL<2800$~\cite{Planck-NG}, we obtain 
a model-independent upper bound on
\eq{|A|<1.2\times10^{-3}\frac{|\Delta\phi_0|}{\sqrt{P_{\delta\phi_{0*}}}}.}
To achieve the currently observed value $|A|\sim0.07$, we would typically need
\eq{\frac{|\Delta\phi_0|}{\sqrt{P_{\delta\phi_{0*}}}}>56\,.\label{eq-M-p1}}

\secs{Simple scenarios}
Here we discuss the possibility of creating the asymmetry $A$ by either a curvaton~\cite{david,enqvist,moroi}, or a spectator field~\cite{Mazumdar:2012rs}, which we denote as $\phi_0$. We may assume that the curvaton is fully embedded within a visible sector not to excite dark radiation~\cite{curv-vis}. The effective potential during inflation can be written as
\eq{V\subs{tot}=V(\phi_i)+U(\phi_0).}

If the perturbation $\Delta\phi_0$ comes directly from the initial fluctuations of $\phi_0$ at some scale $k=a_sH_s$, where we have used the subscript $s$ for the Hubble exit of the initial fluctuations, this perturbation has the order
\eq{\Delta\phi_{0s}\sim\frac{H_skr_{ls}}{\pi}.}
Here $r_{ls}$ is our distance to the last scattering surface.

After the Hubble exit, the initial fluctuation $\Delta\phi_0$ evolves according to
\eq{\frac{\Delta\phi_{0s}}{U_s'}=\frac{\Delta\phi_0}{U'}.}
At the Hubble exit corresponding the pivot scale, indicated by the subscript ``$*$'', we will have
\eq{\frac{|\Delta\phi_0|}{\sqrt{P_{\delta\phi_{0*}}}}\sim B\frac{U_*'}{U_s'}\frac{a_s}{a_*}\frac{H_s^2}{H_*^2}.\label{eq-R-r1}}
where $B=2a_*H_*r_{ls}\sim{\cal O}(1)$, and for inflation we can approximate $H_s/H_*\sim1$. Also, we have
\eqa{\frac{U_*'}{U_s'}&=&e^{-\int_{N_*}^{N_s}\eta_{\phi_0}(N)\dd N},\\
\frac{a_s}{a_*}&=&e^{N_*-N_s}=e^{-\int_{N_*}^{N_s}\dd N}.}
Since $\eta_{\phi_0}>-1$ is the second order slow roll parameter for $\phi_0$, substituting the above relations into \refeq{R-r1}, we find
\eq{\frac{|\Delta\phi_0|}{\sqrt{P_{\delta\phi_{0*}}}}\sim Be^{-\int_{N_*}^{N_s}(1+\eta_{\phi_0}(N))\dd N}<B.\label{eq-R-p2}}

Therefore we can see that as long as the second order slow roll condition is satisfied, $\eta_{\phi_0}>-1$, no matter how its potential varies it cannot overcome the redshift effect during inflation. In this regard our conclusions are similar to the case of Ref.~\cite{Dai:2013kfa}, but our results are more generic.

\section{Speculative ideas}
There are number of ways \refeq{M-p1} could be satisfied. Here we propose a few possibilities.
\begin{itemize}
\item In footnote \ref{fn-multi}, we have argued that if the initial fluctuations of many fields contribute collectively to $|A|$, then they can bring about a larger $\Delta\phi$. Therefore \refeq{M-p1} can be easily satisfied if there are many such fields. One such possibility could arise from assisted inflation~\cite{Assist}.

However it is a fine tuned scenario to generate and coordinate these perturbations, so their effects on $|A|$ are added up instead of canceling out.
\item When the second order slow roll condition is violated, even if very violently for a very short time, it may overcome the exponential suppression in \refeq{R-p2}.
\item As noticed earlier, there are other sources of creating large $-\fNL$ in the context of preheating, which does not rely on initial fluctuations during slow-roll inflation in previous sections, for a review on preheating, see~\cite{preheat}. During preheating, the non-local terms can become important, due to the mode-mode coupling. Inherently, this will gives rise to large non-Gaussianity~\cite{Enqvist:2004ey,Jokinen:2005by}.

One can now imagine that there could be a fine cancellation of various different sources of bispectrum, which might lead to a small $\fNL$ but a large anisotropy $A$.
\end{itemize}

\section{Conclusion}
In this paper we have provided a general formula for more than one sources independently seeding large positive and negative 
$\fNL$, but the resultant non-Gaussianity remains tiny in order to match the observed data~\cite{Planck-NG}. In principle this {\it fine-tuned} cancellation can account for the hemispherical anisotropy observed in the CMB~\cite{Planck-anisotropy}. Although we have found that a simple example of uncoupled fields -- inflaton or curvaton /spectator can lead to $+\fNL$ and $-\fNL$, they cannot give rise to the observed large dipole asymmetry. On the other hand, during preheating large fluctuations can be created, which can account for the dipole asymmetry. This latter scenario presents an interesting possibility which deserves more systematic study.

{\it Acknowledgement:} AM would like to thank Marc Kamionkowski, Takeshi Kobayashi, David Lyth and Tarun Souradeep for helpful discussions. LW would like to thank Youhua Xu for helpful discussions. AM is supported by the Lancaster-Manchester-Sheffield Consortium for Fundamental Physics under STFC grant ST/J000418/1.


\end{document}